\pdfoutput=1
\documentclass[iop]{emulateapj}  
\usepackage{amsmath}
\usepackage{graphicx}
\usepackage{txfonts}
\usepackage{hyperref}
\usepackage{rotating}

\DeclareFontFamily{U}{euc}{}
\DeclareFontShape{U}{euc}{m}{n}{<-6>eurm5<6-8>eurm7<8->eurm10}{}%
\DeclareSymbolFont{AMSc}{U}{euc}{m}{n} 
\DeclareMathSymbol{\umu}{\mathord}{AMSc}{"16}

\slugcomment{\quad}

\shorttitle{Resolving the planet-hosting inner regions of
    the LkCa\,15 disk}
\shortauthors{Thalmann et al.}

\begin{document} 

    
\title{\vspace*{-10mm}Resolving the planet-hosting inner regions of the
    L\lowercase{k}C\lowercase{a}\,15 disk\altaffilmark{$\star$}} 
    
  \author{C. Thalmann\altaffilmark{1},    
          M. Janson\altaffilmark{2},
          A. Garufi\altaffilmark{1,3,4},
          A. Boccaletti\altaffilmark{5},
          S.P. Quanz\altaffilmark{1},
          E. Sissa\altaffilmark{6,7},
          R. Gratton\altaffilmark{6},
          G. Salter\altaffilmark{8}, 
          M. Benisty\altaffilmark{9,10},
          M. Bonnefoy\altaffilmark{9,10},
          G. Chauvin\altaffilmark{9,10},
          S. Daemgen\altaffilmark{1},
          S. Desidera\altaffilmark{6},
          C. Dominik\altaffilmark{11},
          N. Engler\altaffilmark{1},
          M. Feldt\altaffilmark{12},
          T. Henning\altaffilmark{12},
          A.-M. Lagrange\altaffilmark{9,10},
          M. Langlois\altaffilmark{13,8},
          J. Lannier\altaffilmark{9,10},
          H. Le Coroller\altaffilmark{8},
          R. Ligi\altaffilmark{8},
          F. M\'enard\altaffilmark{9,10},
          D. Mesa\altaffilmark{6},
          M.R. Meyer\altaffilmark{1},
          G.D. Mulders\altaffilmark{14,15},
          J. Olofsson\altaffilmark{12,16,18},
          C. Pinte\altaffilmark{19,20,9,10},
          H.M. Schmid\altaffilmark{1},
          A. Vigan\altaffilmark{8,18},
          A. Zurlo\altaffilmark{21,22,8}
          \vspace*{-3mm}
    }

    \altaffiltext{$\star$}{Based on data collected at the European Southern 
        Observatory, Chile (ESO Programme 096.C-0248(A)).}
    \altaffiltext{1}{ETH Zurich, Institute for Astronomy, Wolfgang-Pauli-Strasse 27, 
   		8093 Zurich, Switzerland; \email{thalmann@phys.ethz.ch}}
    \altaffiltext{2}{Department of Astronomy, Stockholm University, 
        106 91, Stockholm, Sweden}
    \altaffiltext{3}{Universidad Auton\'{o}noma de Madrid, Dpto.\ F\'{i}sica Te\'{o}rica, M\'{o}dulo 15, Facultad de Ciencias, Campus de Cantoblanco, E-28049 Madrid, Spain}
    \altaffiltext{4}{Unidad Asociada CAB-UAM}
    \altaffiltext{5}{LESIA, Observatoire de Paris, PSL Research Univ., CNRS, Univ.\ Paris Diderot, Sorbonne Paris Cit\'e, UPMC Paris 6, Sorbonne Univ., 5 place Jules Janssen, 92195 Meudon CEDEX, France}
    \altaffiltext{6}{INAF -- Osservatorio Astronomico di Padova, Vicolo dell'Osservatorio 5, 35122 Padova, Italy}
    \altaffiltext{7}{Dipartimento di Fisica e Astronomia ``G.\ Galilei'' -- Universit\`a degli Studi di Padova, Vicolo dell'Osservatorio 3, 35122 Padova, Italy}
    \altaffiltext{8}{Aix Marseille Universit\'e, CNRS, LAM (Laboratoire d'Astrophysique de Marseille) UMR 7326, 13388, Marseille, France}
    \altaffiltext{9}{Univ.\ Grenoble Alpes, IPAG, F-38000 Grenoble, France}
    \altaffiltext{10}{CNRS, IPAG, F-38000 Grenoble, France}
    \altaffiltext{11}{Anton Pannekoek Institute, University of Amsterdam, Science Park 904, 1098 XH Amsterdam, The Netherlands}
    \altaffiltext{12}{Max-Planck-Institut f\"ur Astronomie, K\"onigstuhl 17, 69117 Heidelberg, Germany}
    \altaffiltext{13}{CRAL, UMR 5574, CNRS, Universit\'e Lyon 1, 9 avenue Charles Andr\'e, 69561, Saint Genis Laval Cedex, France}
    \altaffiltext{14}{Lunar and Planetary Laboratory, The University of Arizona, Tucson, AZ 85721, USA}
    \altaffiltext{15}{Earths in Other Solar Systems Team, NASA Nexus for Exoplanet System Science}
    \altaffiltext{16}{Instituto de F\'isica y Astronom\'ia, Facultad de Ciencias, Univ.\ de Valpara\'iso, Av.\ Gran Breta\~na 1111, Playa Ancha, Valpara\'iso, Chile}
    \altaffiltext{17}{ICM nucleus on protoplanetary disks, Univ.\ de Valpara\'iso, Av.\ Gran Breta\~na 1111, Valpara\'iso, Chile}
	\altaffiltext{18}{European Southern Observatory, Alonso de Cordova 3107, Vitacura, Santiago, Chile}
	\altaffiltext{19}{UMI-FCA, CNRS/INSU, France (UMI 3386)}		
	\altaffiltext{20}{Dept.\ de Astronom\'{\i}a, Universidad de Chile, Santiago, Chile}
    \altaffiltext{21}{N\'ucleo  de  Astronom\'ia,  Facultad  de  Ingenier\'ia,  Universidad  Diego Portales, Av.\ Ejercito 441, Santiago, Chile}
    \altaffiltext{22}{Millenium Nucleus ``Protoplanetary Disk'',  
    Departamento de Astronom\'ia, Universidad de Chile, Casilla 36-D, 
    Santiago, Chile}

 
  \begin{abstract}
%
LkCa~15 hosts a pre-transitional disk as well as at least one 
accreting protoplanet orbiting in its gap. Previous disk observations have focused mainly on the outer disk, which is cleared inward of $\sim$50\,au. The planet candidates, on the other hand, reside at orbital radii around 15\,au, where disk observations have been unreliable until recently. 
Here we present new $J$-band imaging polarimetry of LkCa~15 with SPHERE IRDIS, yielding the most accurate and detailed scattered-light images of the disk to date down to the planet-hosting inner regions.  We find what appear to be persistent asymmetric structures in the scattering material at the location of the planet candidates, which could be responsible at least for parts of the signals measured with sparse-aperture masking.
These images further allow us to trace the gap edge in scattered light at all position angles and search the inner and outer disks for morphological  substructure. 
The outer disk appears smooth with slight azimuthal variations in polarized surface brightness, which may be due to shadowing from the inner disk or a two-peaked polarized phase function. We find that the near-side gap edge revealed by polarimetry matches the sharp crescent seen in previous ADI imaging very well. Finally, the ratio of polarized disk to stellar flux is more than six times larger in $J$-band than in the $RI$ bands.
 \end{abstract}



   \maketitle
   
%



\section{Introduction}

Transitional disks have properties that can be interpreted as intermediate between the protoplanetary and debris disk phases, with an optically thick outer disk and typically a wide (tens of au) gap within \citep[e.g.,][]{strom1989,calvet2005}. 
One possible origin for such gaps is dynamical clearing by forming planets.
The transitional disk host LkCa~15
has emerged as a benchmark target for this scenario.
LkCa~15 is a K5-type star at a distance of $\sim$140\,pc located in the Taurus-Auriga star-forming region \citep{simon2000}. Resolved imaging ranging from the visible to millimeter wavelengths \citep[e.g.,][]{andrews2011,thalmann14,isella14} confirmed the morphological structure implied from spectral energy distribution (SED) fitting \citep{espaillat2007}, all showing 
an outer disk truncated inward of $\sim$50~au.
Recent scattered light imaging 
also revealed 
an inner disk component within the gap structure \citep{thalmann15,oh16}. 

LkCa~15 stands out in that directly imaged planet candidates have been reported in the system. The first candidate was published by \citet{kraus12}, and subsequently two more 
were proposed by \citet{sallum15}. We follow the three-object nomenclature of \citet{sallum15}, in which `b' refers to a candidate at $\sim$14.7~au which displays H$\alpha$ emission, with 
`c' and `d' following counterclockwise.
The H$\alpha$ emission of `b' makes a particularly compelling case for a protoplanet, as it implies active gas accretion. 

Here we present new deep, high-contrast images of LkCa~15 with SPHERE 
(Spectro-Polarimetric High-contrast Exoplanet REsearch; \citealt{SPHERE}),
offering the first full map of the gap's outer edge and revealing the inner disk structures hinted at by earlier studies.
These images also serve to complement the observations in which the planet candidates were discovered, where the precise morphology of the emission cannot be uniquely determined (with the exception of the H$\alpha$ imaging), since they are based on sparse-aperture masking (SAM) and thus do not fully sample the pupil plane of the telescope. 



\section{Observations and data reduction}
\label{s:obs}
On 2015-12-19, we obtained two sets of imaging
polarimetry data with the IRDIS branch \citep{IRDIS, IRDIS-PDI} of
SPHERE at the European Southern Observatory's
Very Large Telescope.  The extreme adaptive optics
(SAXO; \citealt{SAXO}) yielded a point-spread function (PSF) with 
a FWHM of $\sim$4 pixels ($\approx50$\,mas) in the broadband $J$
filter (\texttt{BB\_J}, 1.26\,$\mu{}$m) under good conditions
(seeing 0\farcs6--0\farcs8 in $V$-band).

The first observation, hereafter 
named \textsc{Deep}, aimed at maximizing sensitivity for the faint 
disk structures and therefore employed the \texttt{N\_ALC\_YJ\_S}
coronagraph (apodized Lyot, inner working angle 80\,mas) with long
exposures (32\,s). The observation comprised
15 polarimetric cycles of 2 exposures at each of the 4 standard
half-wave plate positions for a total of 3840\,s of exposure.

The second observation, \textsc{Fast}, was executed immediately
afterwards without 
coronagraph and with minimal exposure times (0.85\,s) to leave the
star unsaturated and grant access to the innermost regions.  Three 
polarimetric cycles with 40 exposures per position yielded
102\,s of exposure.

Frame registration was performed on \textsc{Fast} by 
fitting two-dimensional Gaussians to the unsaturated target star in
each frame.  Since those measured star locations appeared stable 
(jitter $\sim$2\,mas) with no visible 
drift, we adopted their median star location for all \textsc{Deep}
frames. Apart from removing the coronagraph, no changes were made
to the instrument setup between \textsc{Deep} and \textsc{Fast} runs.

The data were reduced using custom \texttt{IDL} routines implementing
the double-ratio polarimetry and empirical correction of instrumental
and stellar polarization from \citet{avenhaus14}.  
We present the results as
polar-coordinate Stokes components ($Q_\phi$, $U_\phi$), in which 
positive $Q_\phi$ signals correspond to azimuthally oriented linear
polarization \citep{quanz13, avenhaus14, thalmann15}. In face-on or mildly inclined circumstellar disks, this convention maps almost all scattered light into positive $Q_\phi$ signal. 
However, due to LkCa~15's inclination of 50$^\circ$ \citep{thalmann14}, multiple-scattering events also produce faint $U_\phi$ signals \citep{canovas15}. $U_\phi$ therefore provides an upper limit for the noise level.

Furthermore, on 2015-11-29, we obtained non-polarimetric
imaging, hereafter named \textsc{Full}, with SPHERE IRDIS 
in dual-band imaging mode (DBI; \citealt{vigan10}) with 
the \texttt{K1K2} filter
($\lambda_{K1} = 2.11$ $\mu$m, $\lambda_{K2} = 2.25$ $\mu$m) as 
part of the SpHere INfrared survey for Exoplanets (SHINE; 
Chauvin et al.\ 2016 in prep).
Pupil tracking was used to 
allow for angular differential imaging (ADI, \citealt{marois06}). We obtained a sequence of 4288\,s totalling a field rotation of 24.5$^\circ$, using the \texttt{N\_ALC\_YJH\_S} coronagraph (apodized Lyot, inner working angle 93\,mas). Non-coronagraphic frames were obtained before and after the coronagraphic sequence for photometric calibration. All data were reduced with the SPHERE pipeline \citep{pavlov08} and additional analysis tools implemented at the SPHERE data centre,
including ADI with \texttt{KLIP} \citep[5 subtracted modes, 24.5$^\circ$ field rotation;][]{soummer12,pueyo15}.
However, since these observations are less 
sensitive to planets than previous studies, we only use them to 
complement our disk analysis (see Section~\ref{s:outer}).


\section{Results and Discussion}
\label{s:results}


\subsection{Imagery}
\label{s:images}

\begin{figure*}[tbp]
\centering
\includegraphics[width=\linewidth]{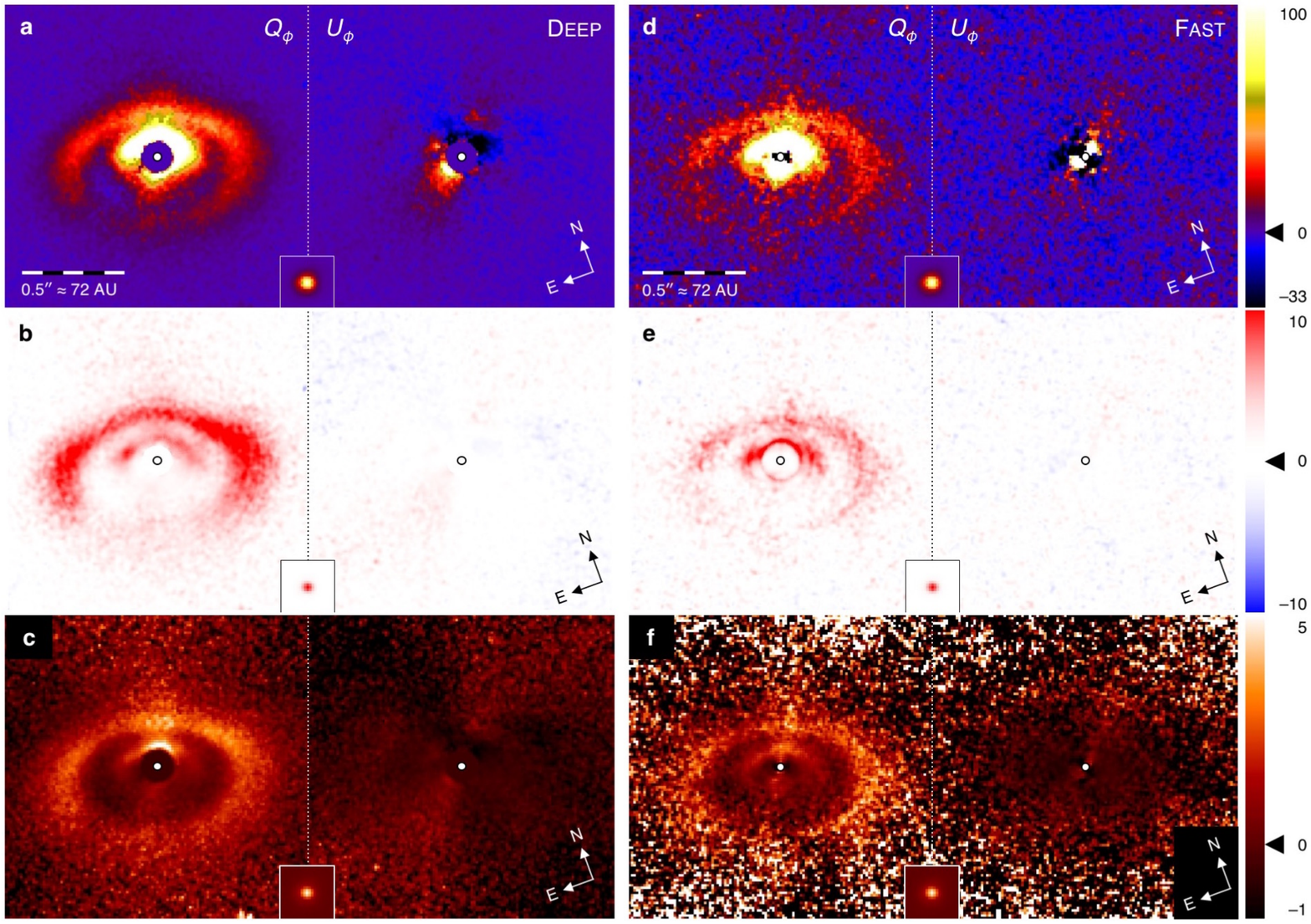}
\vspace*{0.5mm}
\caption{SPHERE IRDIS $J$-band imaging polarimetry of LkCa~15. Each
    panel shows the $Q_\phi$ and $U_\phi$ images side-by-side at the same scale, with insets showing the shape of the PSF core. \textbf{(a)} Polarized flux of \textsc{Deep} at linear stretch (arb.\ units).  The inner disk saturates the color scale. \textbf{(b)} The corresponding S/N map at a stretch of $[-10\sigma,10\sigma]$.  \textbf{(c)} Polarized flux of \textsc{Deep} after scaling with an inclined $r^2$ map to render the faint disk structures visible (arb.\ units). \textbf{(d--f)} The same three images for \textsc{Fast}.  While overall sensitivity is lower in these data, they afford an unobstructed view onto the inner disk. In all panels, the star's location is marked with a white disk. The black wedges on the color scales mark the zero level.}
\label{f:images}
\end{figure*}

Our $J$-band polarimetry of LkCa~15 is shown in
Figure~\ref{f:images}.  The \textsc{Deep} $Q_\phi$ images
(Fig.~\ref{f:images}a) reveal the inner and outer disk components and
the gap between them at much higher S/N than previously 
achieved \citep{thalmann15}.  To quantify the 
detection, we calculate S/N maps (Fig.~\ref{f:images}b) by dividing
the $Q_\phi$ and $U_\phi$ images by a radial noise profile obtained
by calculating the standard deviation of pixel values in concentric
annuli in the $U_\phi$ image.  Both disk components are detected at 
high significance in $Q_\phi$, with local S/N exceeding 5\,$\sigma$
over an area of 0.23\,arcsec$^2$ and peaking at 
11\,$\sigma$.  
Finally, we present the $Q_\phi$ and $U_\phi$ images after scaling
with an $r^2$ map, where $r$ represents the estimated physical 
distance of each pixel's scattering material from the star, assuming it lies on the disk's inclined midplane (50$^\circ$; Fig.~\ref{f:images}c). 
This counteracts the $r^{-2}$ dependency of
the scattered-light intensity and thus reduces the dynamic range of
the image, rendering disk components at different separations directly
comparable.  This reveals the inner edge of the outer disk as a sharp
intensity step ($\sim$60\% over 1\,FWHM, cf.\ Section~\ref{s:profiles})
in $Q_\phi$, allowing us to trace the edge at all 
position angles for the first time.  The residuals in the $U_\phi$ 
image appear as a faint quadrupole pattern diagonal to the principal
axes of the projected disk, as expected from higher-order scattering
\citep{canovas15}. The brightest structures in $r^2$-scaled $U_\phi$
are $\sim$4 times fainter than those in $r^2$-scaled $Q_\phi$.

Figures~\ref{f:images}d--f show the same images for \textsc{Fast}. Despite the lower sensitivity, the same structures as in
\textsc{Deep} can be discerned.  The calculated S/N drops sharply
within a radius of $\sim$80\,mas, though this could be due to 
astrophysical signal from higher-order scattering dominating the
$U_\phi$ image at those separations, causing the noise to be 
overestimated.  For now, we assume that \textsc{Fast} does not
detect the disk at smaller separations than \textsc{Deep}.
However, the disk structures appear notably sharper in 
\textsc{Fast}.  This may be due to the fact that those were 
registered on a frame-to-frame basis, whereas the long exposure
times and global registration of \textsc{Deep} may have 
introduced some smearing.

\begin{figure*}[tbp]
\centering
\includegraphics[width=\linewidth]{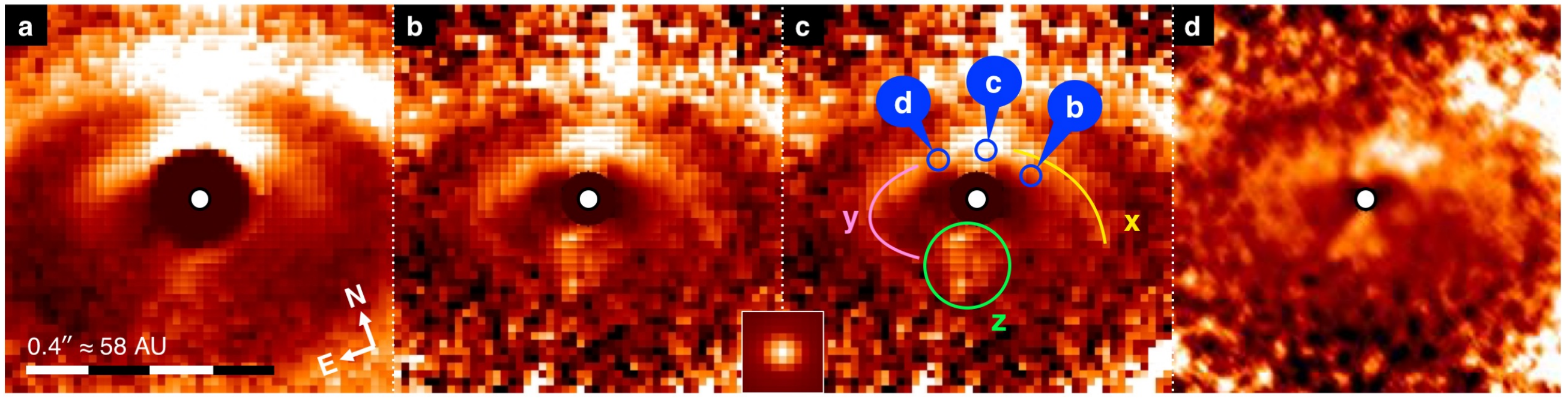}
\caption{Structures in the inner disk of LkCa~15. \textbf{(a)} A
    close-in view of the $r^2$-scaled \textsc{Deep} $Q_\phi$ image. The coronagraph's IWA is 0\farcs08. \textbf{(b)} The same for \textsc{Fast}. The inner disk appears sharper than in (a), perhaps due to better image registration. \textbf{(c)} Image (b) with annotated features. The dark blue circles marked `b, c, d' represent the positions of the three point sources reported in \citet{sallum15}. Source `b' is the one detected in H$\alpha$ imaging.
    The pastel-colored markings identify potential persistent structures in the inner disk: Two curved structures that may represent a disk edge or spiral arms (`x', `y') and a diffuse region along the minor axis (`z'). \textbf{(d)} The visible-light image from SPHERE ZIMPOL \citep{thalmann15} for comparison, which reproduces those structures at least qualitatively. The small inset illustrates the shape of the PSF core.}
\vspace*{2mm}
\label{f:inner}
\end{figure*}


\subsection{Inner disk structure and planet candidates}
\label{s:inner}

Whereas the imagery in \citet{thalmann15} and \citet{oh16} sufficed to
discover and confirm the presence of scattered light from the inner
disk, our new data now allow us to investigate its 
morphology.  Figure~\ref{f:inner} shows a comparison of the inner
regions of the $r^2$-scaled $Q_\phi$ images from \textsc{Deep} (panel 
a) and \textsc{Fast} (b, c) as well as our previously published 
$RI$-band data (d; \citealt{thalmann15}).

All three images show the inner disk in strikingly similar morphology.
The polarized flux is contained in a roughly elliptical area comparable
in shape and orientation to the outer disk, but about half as large in
each dimension.  We note a number of morphological features that appear persistent between the three images, as marked in Figure~\ref{f:inner}c:\
a bright crescent on the near side of the disk, reminiscent of the
full-intensity scattered-light appearance of the outer disk
\citep{thalmann10, thalmann13}; a slight asymmetry between the two 
crescent arms, with the western arm (`x') seemingly trending outward
while the eastern arm (`y') curls inward; and a local brightening
along the far side of the minor axis (`z').  

Furthermore, the area within this crescent appears darkened, giving
the impression of another gap in the inner disk.  However, we caution
that polarimetric efficiency drops in the innermost resolution elements, where the positive/negative pattern in $Q$ and $U$ is too small to be cleanly resolved \citep{avenhaus2014b}.  This could mimic a gap.

Figure~\ref{f:inner}c also indicates the positions of the three
proposed planet candidates 
\citep[`b', `c', `d'; cf.][LBT joint fit results]{sallum15}, 
which coincide with the bright, forward-scattering side of the
inner disk.  The analysis and interpretation of SAM data is 
challenging and potentially even misleading in the presence of 
scattered light from a disk (e.g.,\ \citealt{olofsson13}, \citealt{sallum15}, \citealt{cheetham15}, Caceres et al.\ in prep.); 
thus, the inner disk might account for some or even
all of the SAM signal attributed to protoplanets.
On the other hand, H$\alpha$ imaging makes a strong case for
a protoplanet at `b', which could be confirmed with a re-detection in H$\alpha$ showing orbital motion.  Such a protoplanet could be 
responsible for the substructure in the inner disk.




\subsection{Outer disk structure}
\label{s:outer}

\begin{figure}[tbp] 
\centering
\includegraphics[width=\linewidth]{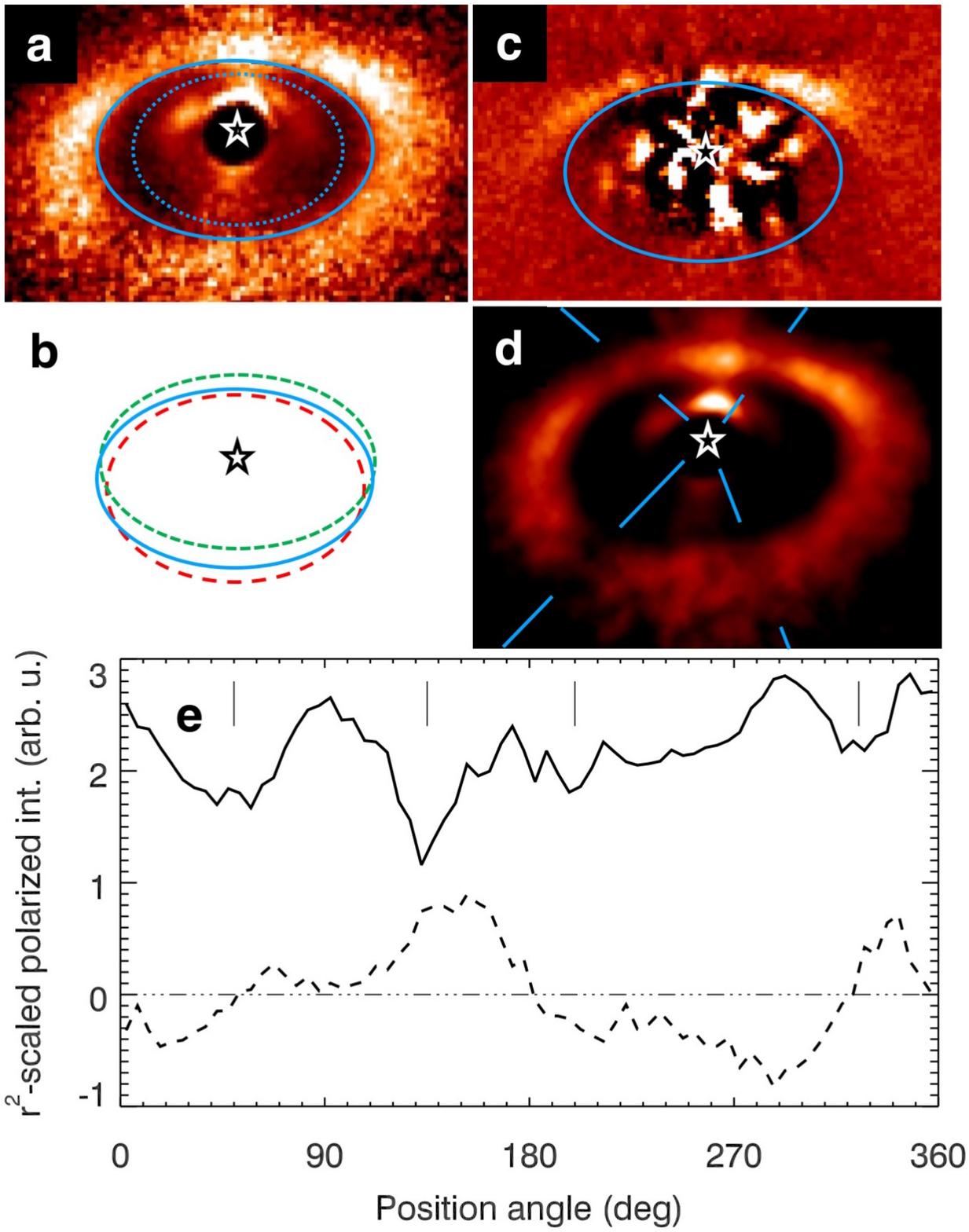}
\vspace*{-1cm}
\caption{Analysis of the outer disk structure of LkCa~15. 
    \textbf{(a)}  Ellipse fits to the maximum gradient (solid blue line) and the flux minimum (dotted blue line) in the $r^2$-scaled \textsc{Deep} $Q_\phi$ image. \textbf{(b)} Comparison of the best-fit gap edge in $J$-band (this work; blue solid line) with those in $RI$-band (\citealt{thalmann15}; red long-dashed line) and sub-millimeter interferometry (\citealt{isella14}; green short-dashed line).  \textbf{(c)} Full-intensity KLIP image (5 subtracted modes) of the \textsc{Full} data
    in the \texttt{K1K2} filter for comparison. The gap edge derived from the $Q_\phi$ image coincides very well with the edge of the bright crescent in the KLIP image.  \textbf{(d)} The image in panel (a) at a harder stretch, emphasizing the surface brightness variations in the outer disk. Four position angles with reduced brightness are marked, possibly indicating transient shadowing from the inner disk. \textbf{(e)} Azimuthal profile of the outer disk in $r^2$-scaled \textsc{Deep} $Q_\phi$ (solid) and $U_\phi$ (dashed),
    evaluated in a 7 pixels (86\,mas) wide annulus outside the best-fit
    gap edge at 5$^\circ$ resolution. 
    The four position angles from panel (d) are indicated with vertical dashes.  The $U_\phi$ profile is dominated by the multiple-scattering quadrupole.}
\label{f:ellipses}
\end{figure}

In the $r^2$-scaled $Q_\phi$ images, the outer disk appears as a 
diffuse region of roughly flat brightness bounded on the inside
by a sharp elliptical edge.  
We employ the maximum merit method \citep{thalmann11} to fit a 
parametric ellipse onto the \textsc{Deep} $Q_\phi$ image that 
maximizes the brightness
contrast between the annular regions immediately inside and 
outside of the ellipse, as detailed in \citet{thalmann15}.  We 
fix the position angle of the major axis to 60$^\circ$ (cf.\
\citealt{thalmann14}). The resulting ellipse is shown in solid
blue in Figure~\ref{f:ellipses}a. The fit ellipse matches the
perceived gap edge accurately.

To separate the inner and outer disk for further analysis, we also
fit a second ellipse along which the $r^2$-scaled $Q_\phi$ intensity
is minimized.
This ellipse is shown in dotted blue in Figure~\ref{f:ellipses}a.
The numerical parameter of both ellipses are given in 
Table~\ref{t:num}. For errors, we ``fit'' similar ellipses to 
$U_\phi$ and measure the standard deviation of the resulting merits.
The error intervals represent the family of ellipses whose $Q_\phi$
merits are less than one standard deviation below the best fit.

Figure~\ref{f:ellipses}b compares the gap edge fit ellipses in 
$J$-band polarimetry (this work; solid blue) to those in $RI$-band
polarimetry (\citealt{thalmann15}; long-dashed red) and 
sub-millimeter interferometry (\citealt{isella14}; short-dashed
green) with respect to the star's position.  
Our new data confirm the eccentric scattered-light gap
contrasting with the symmetric gap in sub-millimeter thermal
emission.  Due to the tapered gap wall, inclination cannot explain
this eccentricity \citep{thalmann14}, whereas shadowing from the
inner disk could \citep{thalmann15}.
The eccentricity appears slightly lower than 
in $RI$-band.  This could be due to the poorly constrained far-side
edge in the $RI$-band data, color-dependent polarized reflectivity, chromatic filtering by the inner disk halo, or a temporal 
evolution of the shadowing postulated as the origin of the eccentric
scattered-light gap.

Figure~\ref{f:ellipses}c shows the $J$-band gap ellipse superimposed
on the $K1K2$-band full-intensity KLIP image of the \textsc{Full} 
data.  The ellipse coincides exactly with the 
sharp inner edge of 
the bright forward-scattering crescent, confirming the long-held
assumption that such crescents in ADI images accurately trace the
near-side gap edge even though the fainter parts of the disk suffer 
from oversubtraction effects.

The \textsc{Deep} data are sensitive enough to search for
substructure in the outer disk. While we find no sign of spirals or structural asymmetries, we note significant variations in the 
azimuthal $Q_\phi$ distribution. In particular, the disk appears
locally dimmed in radial lanes around azimuths
50$^\circ$, 135$^\circ$, 200$^\circ$, and 325$^\circ$ (see Figure~\ref{f:ellipses}d, e). Furthermore, near-side gap edge shows
similar substructure in the \textsc{Full} and \textsc{Deep} images.

One possible scenario is that these dark lanes indicate shadows cast by inner disk regions \citep{Marino2015, Stolker2016} or magnetospheric accretion columns \citep{bodman16}. 
The inner and outer disks do have similar azimuthal structure.

An alternative explanation is that a polarized scattering phase 
function with peaks in both the forward and backward 
directions \citep[e.g.,][]{Min2016} could brighten the projected disk along its minor axis (Stolker et al.\, in prep.).
A case study for this scenario is HD100546 \citep{Garufi2016}, where a bright wedge surrounded by two dark wedges was detected along the near-side minor axis. Another analogy between these two datasets is the presence of a tenuous counterpart to the bright wedge in $U_\phi$, which is roughly 45$^\circ$ displaced from it. This detection may suggest small departures from centrosymmetric scattering as expected from multiple scattering in inclined disks \citep{canovas15}.


\subsection{Polarized surface brightness profiles}
\label{s:profiles}

To quantify the amount of polarized light detected, we calculate
surface brightness profiles in the $Q_\phi$ image without $r^2$ 
scaling.  We extract a 86\,mas wide strip along the major axis of 
the projected disk (position angle 60$^\circ$; \citealt{thalmann14})
and a 257\,mas wide strip along the minor axis,
and average the pixel values across the width to obtain 
one-dimensional profiles.  We then divide each profile by the
sum of all pixel values in the unsaturated stellar full-intensity
PSF (out to 1\farcs23) and by the solid angle on sky
corresponding to a pixel (0\farcs0123$^2$).  The
values of the resulting profiles therefore measure the local 
polarized flux density of the disk in units of stellar flux per 
square arcsecond. The resulting values do not explicitly depend
on pixel size or wavelength and therefore allow for direct 
comparison between optical and infrared measurements.  (This
differs from the normalization used in \citealt{thalmann15}.)

\begin{figure}[tbp] 
\centering
\includegraphics[width=\linewidth]{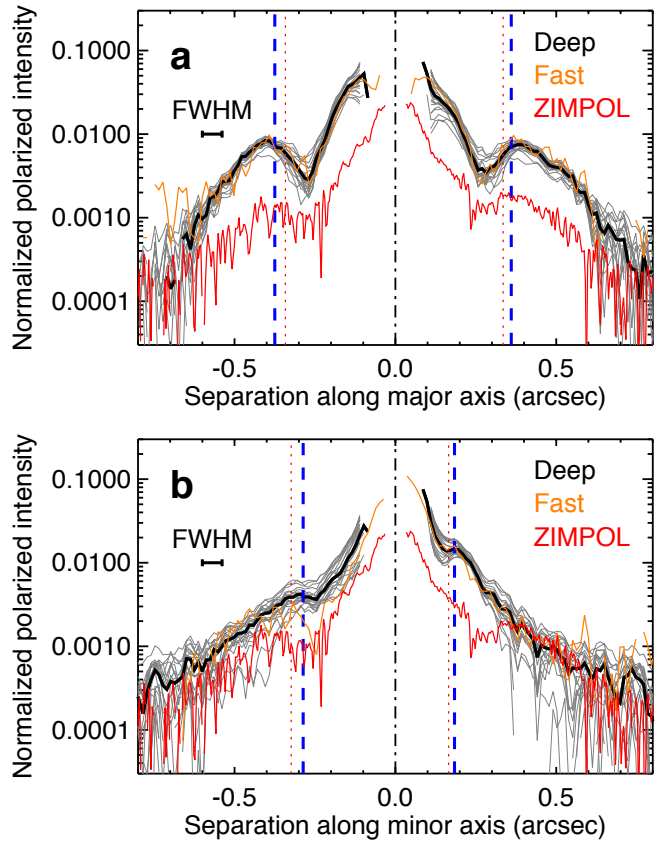}
\caption{Polarized surface brightness profiles of the LkCa~15 disk
    in stellar flux per square arcsecond (see text).
    \textbf{(a)} Profiles measured in a 86\,mas wide strip along the major axis, centered on the star.  Positive
    separations correspond to a position angle of 60$^\circ$ on sky.
    \textbf{(b)} Profiles measured in a 257\,mas wide strip along the minor axis, centered on the star.  Positive separations correspond
    to a position angle of 330$^\circ$.
    Solid black lines show the \textsc{Deep} $J$-band $Q_\phi$ surface brightness profiles. The family of thin grey lines represents the results of adding strips from the $U_\phi$ image taken from different position angles to the $Q_\phi$ strip to illustrate the estimated noise level. Thin orange lines show the corresponding profiles from  \textsc{Fast}, which match \textsc{Deep} very well. Thin red lines represent the $RI$-band profiles from \citet{thalmann15} for comparison, demonstrating that the disk is significantly more reflective in polarization in $J$-band than in $RI$-band. Dashed blue vertical lines demarcate the bounds of the $J$-band gradient fit ellipse from Fig.~\ref{f:ellipses}. Dotted red vertical lines show the corresponding bounds for the $RI$-band fit ellipse from \citet{thalmann15}.}
\label{f:profiles}
\vspace*{5mm}
\end{figure}

Figures~\ref{f:profiles}a and b show the profiles for the major
and minor axis, respectively.  The solid black curve
corresponds to \textsc{Deep}.
To visualize the uncertainty level, we extract a number of 
equivalent profiles from the \textsc{Deep} $U_\phi$ image and add
them to the measured $Q_\phi$ profiles, yielding the family of
grey curves.  The thin orange curves are the $Q_\phi$ profiles
for \textsc{Fast}, which match the \textsc{Deep} results
well.  We also show the corresponding $Q_\phi$ profiles for the
$RI$-band data from \citet{thalmann15} as thin red curves.  While
the morphology of the $RI$- and $J$-band profiles is similar, 
the latter is significantly brighter overall.

We investigate this effect by summing the $Q_\phi$ image over the 
detectable parts of the inner and outer disks in each filter.  
We sample the inner disk
from a radius of 70\,mas out to the flux-minimum fit ellipse
(dotted blue in Fig.~\ref{f:ellipses}a), and the outer disk 
from that ellipse out to a radius of 1\farcs225.  We find that
the polarized disk flux to stellar flux ratio is 6.4 times 
larger in $J$ than in $RI$ for the inner disk, and 
6.6 times for the outer disk.  The polarized flux ratio between
the inner and outer disk area appears stable with wavelength 
(0.70 in $J$, 0.73 in $RI$).  Judging from the $U_\phi$ images,
these figures are accurate on the order of 10\%.  The numerical
results are summarized in Table~\ref{t:num}.

\begin{table}[tbp]
\caption{Numerical results.}
\label{t:num}
\footnotesize 
\begin{tabular}{l|r@{~~}r@{~}r@{\qquad}r@{~~}r@{~}r@{}}
\multicolumn{1}{@{}l}{} &   \multicolumn{3}{c}{$J$-band}    &
        \multicolumn{3}{c}{$RI$-band}\\
    \hline
\multicolumn{1}{@{}l|}{\emph{Outer disk gap edge}} \\
Semimajor axis $a$ (mas)    & 367 & $(+15,$ & $-25)$ 
    & 338 & $(+11,$ & $-18)$ \\
\qquad --- (au)             & 51.5 & $(+2.1,$ & $-3.4)$ 
    & 47.4 & $(+1.5,$ & $-2.5)$ \\
Semiminor axis $b$ (mas)    & 235 & $(+10,$ & $-15)$
    & 245 & $(+21,$ & $-11)$ \\
\qquad --- (au)             & 32.9 & $(+1.4,$ & $-2.1)$   
    & 34.3 & $(+3.0,$ & $-1.5)$ \\
Major-axis offset $x$ (mas) & $-7$ & $(+20,$ & $-15)$
    & $-3$ & $(+11,$ & $-14)$ \\
\qquad --- (au)             & $-1.0$ & $(+2.7,$ & $-2.1)$
    & $-0.5$ & $(+1.5,$ & $-2.0)$ \\
Minor-axis offset $y$ (mas) & $-51$ & $(+15,$ & $-10)$
    & $-79$ & $(+11,$ & $-18)$ \\
\qquad --- (au)             & $-7.2$ & $(+2.1,$ & $-1.4)$
    & $-11.1$ & $(+1.5,$ & $-2.5)$ \\[2mm]
\multicolumn{1}{@{}l|}{\emph{Gap minimum}} \\
Semimajor axis $a$ (mas)    & 276 & $(+92,$ & $-104)$ \\
\qquad --- (au)             & 38.6 & $(+12.9,$ & $-14.6)$ \\
Semiminor axis $b$ (mas)    & 196 & $(+0,$ & $-76)$ \\
\qquad --- (au)             & 27.4 & $(+0,$ & $-10.6)$ \\
Major-axis offset $x$ (mas) & $0$ & $(+12,$ & $-37)$ \\
\qquad --- (au)             & $0$ & $(+1.7,$ & $-5.1)$ \\
Minor-axis offset $y$ (mas) & $-51$ & $(+27,$ & $-22)$ \\
\qquad --- (au)             & $-7.2$ & $(+3.8,$ & $-3.1)$ \\[2mm]
\multicolumn{1}{@{}l|}{\emph{Polarized disk/star contrast}} \\
Inner disk  &\multicolumn{3}{r@{\qquad}}{0.0025}
    &\multicolumn{3}{r@{}}{0.00040} \\
Outer disk  &\multicolumn{3}{r@{\qquad}}{0.0036}
    &\multicolumn{3}{r@{}}{0.00055} \\
Inner/outer ratio & \multicolumn{3}{r@{\qquad}}{0.70}
    &\multicolumn{3}{r@{}}{0.73} \\
\hline
\end{tabular}
\\[2mm]
\textsc{Notes.} $J$-band results are based on this work; $RI$-band
results are taken from \citet{thalmann15}.  The gap fitting parameters
are explained in Section~\ref{s:outer}, the contrast values in 
Section~\ref{s:profiles}. The accuracy of the contrast values is 
estimated at $\sim$10\%.\\
\end{table}

The photometry is likely
affected by PSF-dependent loss of polarimetric sensitivity at 
small separations \citep{avenhaus14}.
However, the fact that we measure consistent color ratios for the
inner and outer disks indicates that the differential effect between
the two bands is small, most likely because both
data sets were taken with the same telescope and AO system.

A higher polarized disk flux at longer wavelengths was also
observed for the transition disk HD~135344~B by \citet{Stolker2016},
who measured a factor $\sim$2 between $J$ and $RI$ bands.
This color factor offers an
additional diagnostic for the composition of scattering dust.
Our ongoing SPHERE observations are measuring polarized colors
for many protoplanetary disks, which will allow for a comparative
study in the near future.


\begin{acknowledgements}
This work has been carried out within the frame of the NCCR PlanetS 
supported by the SNSF, and is supported by ANR-14-CE33-0018.
ETH members acknowledge support from SNSF; INAF members from MIUR's ``Progetti Premiali''; J.O.\ from the Millennium Nucleus RC130007 (Chilean Ministry of Economy) and ALMA/Conicyt Project 31130027.
We thank the referee for constructive input.
%
\end{acknowledgements}

\textit{Facilities:} \facility{VLT:Melipal (SPHERE IRDIS)}



\bibliographystyle{aa}

%

\clearpage

\end{document}